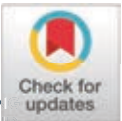

# JGR Space Physics

# A Study of Automatically Detected Flow Channels in the Polar Cap Ionosphere

K. Herlingshaw[1,2], L. J. Baddeley[1,2], K. Oksavik[1,2], D. A. Lorentzen[1,2], and E. C. Bland[1,2]

[1]Department of Arctic Geophysics, University Centre in Svalbard, Longyearbyen, Norway, [2]Birkeland Centre for Space Science, University of Bergen, Bergen, Norway

**Abstract** This paper presents a new algorithm for detecting high-speed flow channels in the polar cap. The algorithm was applied to Super Dual Auroral Radar Network data, specifically to data from the new Longyearbyen radar. This radar is located at 78.2°N, 16.0°E geographical coordinates looking north-east, and is therefore at an ideal location to measure flow channels in the high-latitude polar cap. The algorithm detected >500 events over 1 year of observations, and within this paper two case studies are considered in more detail. A flow channel on "old-open field lines" located on the dawn flank was directly driven under quiet conditions over 13 min. This flow channel contributed to a significant fraction (60%) of the cross polar cap potential and was located on the edge of a polar cap arc. Another case study follows the development of a flow channel on newly opened field lines within the cusp. This flow channel is a spontaneously driven event forming under strong solar wind driving and is intermittently excited over the course of almost an hour. As they provide a high fraction of the cross polar cap potential, these small-scale structures are vital for understanding the transport of magnetic flux over the polar cap.

## 1. Introduction

For southward interplanetary magnetic field (IMF), the high-latitude plasma convection in the ionosphere can be described on the large scale as twin-cell convection, which flows antisunward across the polar cap and returns to the dayside at auroral latitudes. The flow is usually assumed to be constant over thousands of kilometers with typical speeds of several hundreds of meters per second (MacDougall & Jayachandran, 2001). However, on a smaller scale (100–500 km), convection within the polar cap is not a uniform, laminar flow but is instead frequently driven by dynamic mesoscale phenomena. These structured flow enhancements occur at many locations and are classified under different names within the literature depending on their location, speed, size, and duration. Pinnock et al. (1993) reported a class of longitudinally extended flow channel events (FCEs) within Southern Hemisphere radar data, where the plasma within the flow channels traveled with speeds of 2–3 km/s within the dayside cusp. These features were suggested to be an ionospheric manifestation of flux transfer events (FTEs), which are reconnection events at the dayside magnetopause with a mean recurrence rate of 7–8 min (Haerendel et al., 1978; Russell & Elphic, 1978, 1979). The ionospheric response to transient dayside reconnection has been observed and named differently depending on the context in which it was observed. This can introduce confusion as the observations are linked and in some cases can describe the same phenomenon.

One of the other radar signatures of dayside reconnection linked to FTEs and FCEs was detected in Super Dual Auroral Radar Network (SuperDARN) data and named pulsed ionospheric flows (PIFs; Provan et al., 1998, 1999). These periodic bursts of antisunward convection, detected initially by the SuperDARN radar located in Finland, have a typical recurrence rate of 7–8 min but this can vary between 5 and 12 min. Neudegg et al. (2000) identified a statistical link between FTEs and high speed flows in the polar ionosphere (such as PIFs/FCEs) with over 99% confidence by using Equator-S satellite data and SuperDARN data to observe the effects of FTEs propagating from the magnetosphere to the ionosphere. McWilliams et al. (2000) found the occurrence rates and repetition rates of PIFs to be very similar to that of FTEs and poleward moving auroral forms, the optical manifestations of FTEs (Denig et al., 1993; Milan et al., 1999, 2000; Sandholt et al., 1990, 1993; Thorolfsson et al., 2000).

In addition to poleward moving auroral forms, polar cap arcs (PCAs) are also optical features associated with flow channels. PCAs are caused by precipitating electrons that are often accelerated through a field-aligned potential drop and occur primarily under the influence of a northward IMF and quiet geomagnetic conditions. They are known to be associated with flow shears and therefore flow channels, where the potential drop across the flow channel is typically ~10 kV and can account for ~10–40% of the cross polar cap potential (Zou et al., 2015a, 2015b).

Flow channels (FCs) have also been observed in locations other than the cusp, and their temporal evolution has been described by dividing them into four categories: FC 1, FC 2, FC 3, and FC 4 (Sandholt & Farrugia, 2009). FC 1 and FC 2 are driven by dayside reconnection and occur at different stages of evolution of the opened field lines. Both FC 1 and FC 2 arise from the closure of field-aligned currents via ionospheric Pedersen currents, but they differ as FC 1 occurs on newly opened field lines (time since reconnection <10 min), whereas FC 2 occurs on "old-open field lines," which underwent reconnection 10–30 min earlier (Andalsvik et al., 2011; Sandholt & Farrugia, 2009). PIFs occur on newly opened field lines (corresponding to FC 1) in the cusp, while FC 2 is located immediately poleward of the auroral oval on the dawn and dusk flanks (06–09/15–18 magnetic local time (MLT)). FC 2 are 200–300 km wide channels of enhanced antisunward flow which typically last for 5–10 min. They are attributed to momentum transfer from the high-latitude and flank boundary layers on the downstream side of the cusp via the C1-C2 cusp currents. The C1-C2 cusp currents (Farrugia et al., 2003; Sandholt et al., 2010) form the system responsible for momentum transfer from the high-latitude and flank boundary layers on the downstream side of the cusp on old-open field lines. These channels can either be "directly driven" by southward turnings in the IMF after an appropriate time delay, or "spontaneous" events occurring during stable periods of southward IMF where the channel is intermittently excited (Sandholt et al., 2010). An IMF $B_y$-induced asymmetry in the location of FC 2 is noted by Sandholt and Farrugia (2012), as FC 2 is located mainly on the postnoon/dusk (prenoon/dawn) side of the polar cap for IMF $B_y < 0$ (>0) conditions. Andalsvik et al. (2011) expands on this framework by defining a polar cap flow channel as a latitudinally restricted (a few 100 km) regime of enhanced antisunward convection >1 km/s and studies the dayside and nightside sources of polar cap convection events. FC 3 (premidnight/postmidnight sectors) and FC 4 (linked to streamers) are driven by nightside processes in the magetotail lobes and/or plasma sheet. The statistics of the nightside flow channels have been studied by Gabrielse et al. (2018) and they find that the flow channels are aligned with the large-scale background convection. They also find a postmidnight preference in the polar cap flows, which is similar to the behavior of PCAs.

Previous work involving flow channels has largely been based on data from satellite passes (e.g., Sandholt & Farrugia, 2009), where the flow channel can only be sampled once per orbit and therefore do not describe the temporal evolution of the flow channel. Although SuperDARN data have been used in the past to observe flow channels, studies have focused on either a single beam within a radar scan (Provan et al., 1999) or occasionally referenced SuperDARN global convection maps for a large-scale overview of the flow channel (Andalsvik et al., 2011). Observing a single beam within a radar scan allows for the best temporal resolution, but offers limited information about the structure of the channel. Each cell within the global convection map is a 111 km × 111 km square, which places a lower limit on the size of channels which can be detected. To give further insight into the structure and spacial/temporal evolution of flow channels, we will analyze individual, whole field-of-view (FOV) radar scans from the Longyearbyen SuperDARN radar over a period of 1 year.

The selection of a 0.9 km/s velocity magnitude threshold and the definition of a flow channel is discussed in more details in section 3 of the paper. These, along with additional applied criteria, ensure that only fast, well-defined channels embedded within a region of slower, background convection flow are identified (as shown, e.g., in Figure 1). In this paper, the focus will be on two case studies of flow channels, one on the dawn flank and another within the cusp. The properties, formation, evolution and contribution to the cross polar cap potential will be examined. A future paper will study the statistics of the detected flow channels.

## 2. Instrumentation

The Super Dual Auroral Network (SuperDARN) is a global network of over 30 high-frequency coherent scatter radars designed primarily for studying $F$ region ionospheric plasma (Chisham et al., 2007; Greenwald et al., 1995). During the common mode of operation, each SuperDARN radar in the network steps through a series of azimuthally consecutive beams, separated by ∼3° increments. Each beam is divided into 75 range gates of 45 km resolution. The radars are frequency agile (8–20 MHz) and routinely measure the line-of-sight Doppler velocity, spectral width, and backscattered power from magnetic field-aligned ionospheric irregularities. These decameter-scale irregularities drift at the bulk E × B drift velocity in the F region ionosphere.

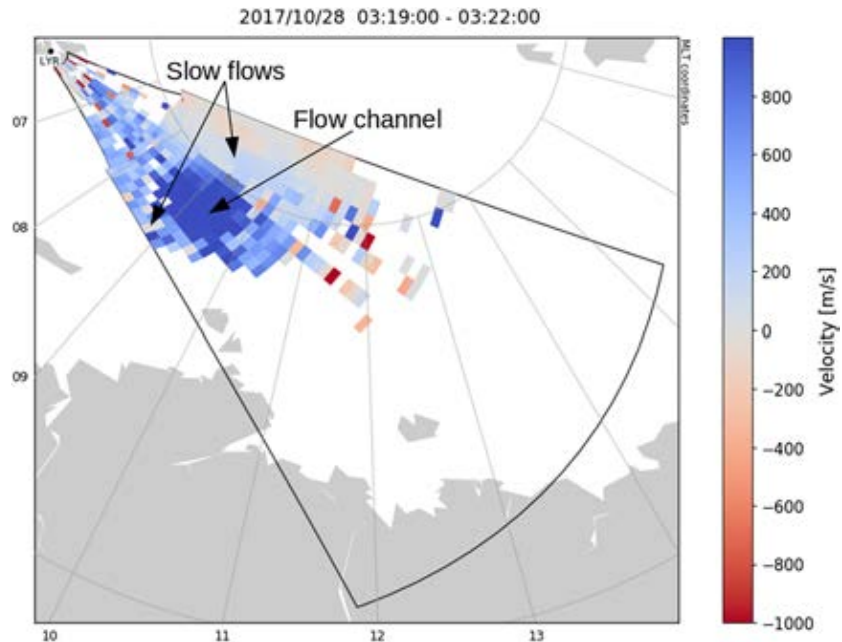

**Figure 1.** The field-of-view of the Longyearbyen radar in Magnetic Local Time (MLT)/Magnetic Latitude (MLat) coordinates showing the line-of-sight velocity, where blue and red represent flows toward and away from the radar, respectively. A clear channel of enhanced flow toward the radar is visibly embedded within the slower moving background flow.

In this study, we specifically used scans from the Longyearbyen SuperDARN (LYR) radar, which is located at 78.153°N, 16.074°E, 472 m altitude and began operations in October 2016. The data used were recorded in common mode in 2017 on channel A (9.8–9.9 MHz) at 1 min resolution, which is the time taken to complete a scan of all 16 beams. The LYR radar was selected as the first target of the flow channel detection algorithm as it has an ideal position at a high-latitude with a north-east facing field-of-view (shown in Figure 1), covering a large area of the polar cap and receiving a large amount of backscatter. The radar can theoretically detect backscatter up to 3,500 km in range but more regularly records data up to 1,500 km with a latitudinal range of approximately 76–82° (magnetic coordinates).

Data from SuperDARN radars can be combined to provide maps of the high-latitude ionospheric convection using the "map potential" technique (Ruohoniemi & Baker, 1998), in which the electrostatic potential pattern is determined as an expansion in spherical harmonic functions. Line-of-sight velocity measurements from all of the radars are gridded and used to determine the values of the spherical harmonic coefficients, while an IMF-driven model is used to constrain the spherical harmonic fit in areas where data coverage is sparse or absent. The model used to generate the convection maps within this study is the TS18 statistical model of ionospheric convection (Thomas & Shepherd, 2018). The electrostatic potential pattern usually reaches a maximum near dawn and a minimum near dusk. The cross polar cap potential (CPCP), a proxy for the strength of the ionospheric convection at a given time, can then be calculated as the difference between the maximum and minimum potential. All magnetic coordinates displayed within this paper are altitude adjusted corrected geomagnetic coordinates (Baker & Wing, 1989; Shepherd, 2014).

This study also uses data from the Defense Meteorological Satellite Program (DMSP). Each DMSP satellite has a 101 min polar, Sun-synchronous orbit at an altitude of 840 km. Since 2003, the DMSP satellites have housed the Special Sensor Ultraviolet Spectrographic Imager (SSUSI), with a global far ultraviolet (UV) imager (165–180 nm; Paxton et al., 1992). Each scan of the oval takes ∼20 min, and the emissions are produced primarily due to precipitating electrons impacting the upper atmosphere. In this paper, observations in the Lyman-Birge-Hopfield Long (LBHL) wavelength range (165–180 nm) are presented to provide a global context for each event.

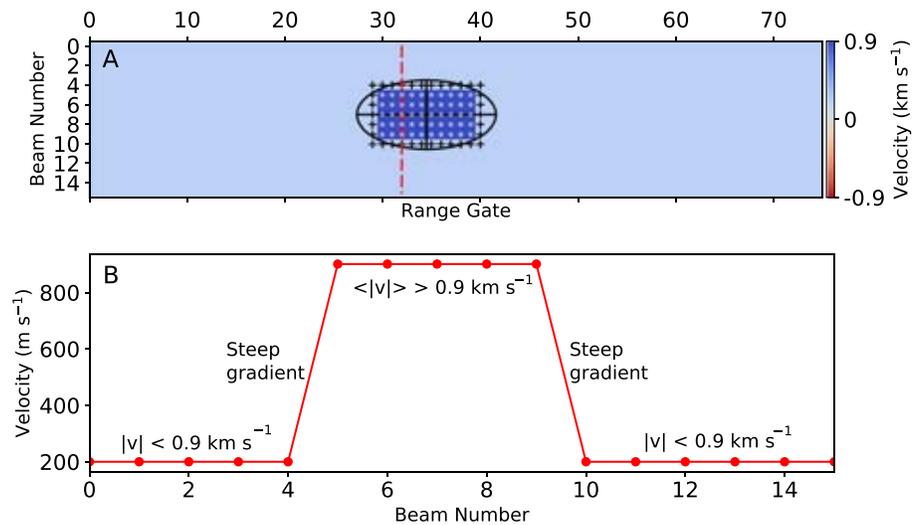

**Figure 2.** Plots demonstrating the analysis of a simulated flow channel. Panel (a) shows a channel of >0.9 km/s flow within slower moving (200 m/s) background flow. Positively identified cells are marked with gray dots, edge cells are marked by black crosses, and the principal component axes are marked with solid black lines and encompassed by an oval. A vertical dotted red line shows a slice through the beams at a constant range. Panel (b) shows the velocity across that slice with annotations of notable features.

Optical images recorded by the Sony a7s All Sky Camera (ASC) at the Kjell Henriksen Observatory (78.148°N, 16.043°E, 520 m altitude) were also used in this study to provide a detailed view of the time evolution of the auroral emissions in the vicinity of the flow channel. The camera is located 1 km from the Longyearbyen radar and uses a fish-eye lens to capture 180° color images of the sky with a 4 s exposure time.

## 3. Flow Channel Detection Algorithm

The aim of the newly developed algorithm was to automatically search through the large SuperDARN data set and locate fast flowing channels embedded within a slower moving background convection. The algorithm was applied on the Longyearbyen radar to all 2017 channel A common mode data. This equated to 314 days of 1 min resolution scans and in total ∼450,000 scans were used. The algorithm requires a larger amount of scatter coverage to be present within each fan plot than is typically available within a single scan. This is due to the fact that the radar relies on backscatter from electron density irregularities where the Bragg condition is satisfied and it is generally not the case that data will be obtained from every range gate along every beam. Therefore, the data set was smoothed to improve data coverage by averaging each individual cell (1 × 45 km range gate along one beam) within the FOV over three consecutive scans. Only the available data in each cell over the three scans were averaged together, so the averages in each cell were calculated from between one and three individual velocity values. From this point onward, we define a "scan" to be a grid of beams and range gates, where each "cell" contains an average velocity at that specific beam and range gate over any available velocity data within three consecutive minutes. A choice of a 3 min average for each scan maintains a high enough resolution to detect transient channels associated with reconnection.

Figure 2a shows a scan of an idealized flow channel plotted as color contours on a beam-gate grid, where the red dotted line represents a slice through the beams at a constant range. This idealized case will be used to illustrate the steps in the algorithm. If there were no fast flows within the grid, then there could be no flow channel, thus a check for velocity magnitudes above 0.9 km/s was applied to the grid. If the search yielded nothing, this grid was neglected, the start time was shifted by 1 min and a new scan was generated. If there were fast flows within the grid, it was important to determine if they contain structure or if they were actually noise and not physical. The cell containing the fast flow was then compared to the eight neighboring cells, spanning across the adjacent beams and ranges. If scatter was present in all of the neighbors, every cell contained velocities of the same sign, and their average velocity exceeded 0.9 km/s, then a velocity structure is said to exist and this represents the threshold of detection of the algorithm.

Another test was required to eliminate instances where fast flows may exist but with a gradual gradient across the structure. Cases such as these are likely to be an artifact of the look direction of the radar, perhaps viewing constant flow at an angle which appears to reveal fast flows gradually transitioning to slower flows. A quantitative check of the gradient is then required and is demonstrated in Figure 2b, which shows the velocity profile over a slice across the beams at a constant range. The gradient of the velocity is calculated by taking the difference in velocity ($v2 - v1$) in a sliding window across the entire slice. The algorithm searches for two sharp gradients of opposing signs, which should be present on the edge cells of the channel at the transition points between the inside of the fast moving channel and the slower background flows. These two locations are defined as the "edge" of the channel and it is now possible to refer to the inside and outside of the channel with respect to these edges. The edge cells must be of the same sign as the inside of the channel, and there must not be any missing data within the channel. The threshold placed on the gradient is 400 m s$^{-1}$ cell$^{-1}$, as this eliminated most of the slowly varying gradients. Additionally, the inside of the channel is examined to ensure that on average the flow magnitude is above 0.9 km/s, which allows for variation but prevents the detection of more complicated channels that are harder to analyze. There should also be no velocity values exceeding 0.9 km/s outside of the channel as this region should be the slower background flow. To ensure that the background flow exists around the channel, the points immediately outside of the edge cells along the slice are examined. If >80% of these points are present and there are >5 edge cells in total, then the FC is accepted to be embedded within a discernible background flow. Slicing only at constant range through the identified cell may also bias results to certain orientations of flow channel. To avoid this, slices are taken every 30° in a circle around the identified cell and the structure was accepted as a flow channel if any one of these slices satisfied the listed criteria of the gradient and background tests.

Another feature shown in Figure 2a is the black oval encompassing the flow channel, where the semimajor and semiminor axes (marked with black straight lines) demarcate the principal component axis of all the identified cells of the algorithm (gray dots). The principal components of the identified cells were calculated to estimate the orientation of the channel with respect to the beam direction (semimajor axis) and the width of the channel (semiminor axis). The ellipse that bounds the flow channel contains 96% of all identified cells.

While a variety of velocity magnitudes have been observed inside flow channels (e.g., Andalsvik et al., 2011, $v > 1$ km/s; Nishimura et al., 2014, $v = 0.9$ km/s; Oksavik et al., 2005, $v = 0.5$ km/s) there exists no definition as to how to define the edge of a channel. As such, we tested the detection algorithm with three different velocity gradient thresholds (400, 500, and 600 m s$^{-1}$ cell$^{-1}$) in combination with a variety of velocity magnitude thresholds inside the channel (from 500 up to 1,000 m/s). To ensure we have fast flows we set the velocity threshold toward the higher end of the range of velocities observed in a flow channel at $v > 0.9$ km/s. Using this as a velocity threshold and given the spatial resolution of the radar data (45 km range gates along the beam), it was decided that a velocity gradient threshold of 400 m s$^{-1}$ cell$^{-1}$ would ensure that we are indeed observing channel structures with distinct edges embedded within a slower background flow. Using this criteria, the flow surrounding the channel would then be a maximum of 45% of the flow inside the channel and this change would occur over a small spatial scale (1 cell). In addition, an examination of the actual velocity inside the identified channels using this criteria was also undertaken, which showed an average channel velocity of ~1,100 m/s with a lower limit of the full width half maximum of the distribution at 950 m/s.

Figure 3 shows four examples of detected channels, where the gray dots indicate the cells which passed all the tests. Many different sizes and orientations of flow channels were identified, and in many cases the flow channel persisted for multiple consecutive scans. To further investigate the location and duration of the flow channels, it was necessary to consider all of the positively identified cells and define a single channel center within each scan. In addition, it is required to separate all of the detections into discrete events by defining a maximum allowable time gap between scans, which if exceeded would separate one "event" from the next. An average of the beams and range gates of the positively identified cells within each scan produced the coordinates of a single location for the flow channel center.

The algorithm identified 546 events with FC centers over a range of MLT as shown in Figure 4. An event is classified as a continuous detection of a FC with no time gaps larger than 3 min. The majority of flow channels were detected in the dayside polar cap between 9 and 14 MLT. There are also events present along the dawn and dusk edges of the polar cap and a small number of events on the nightside.

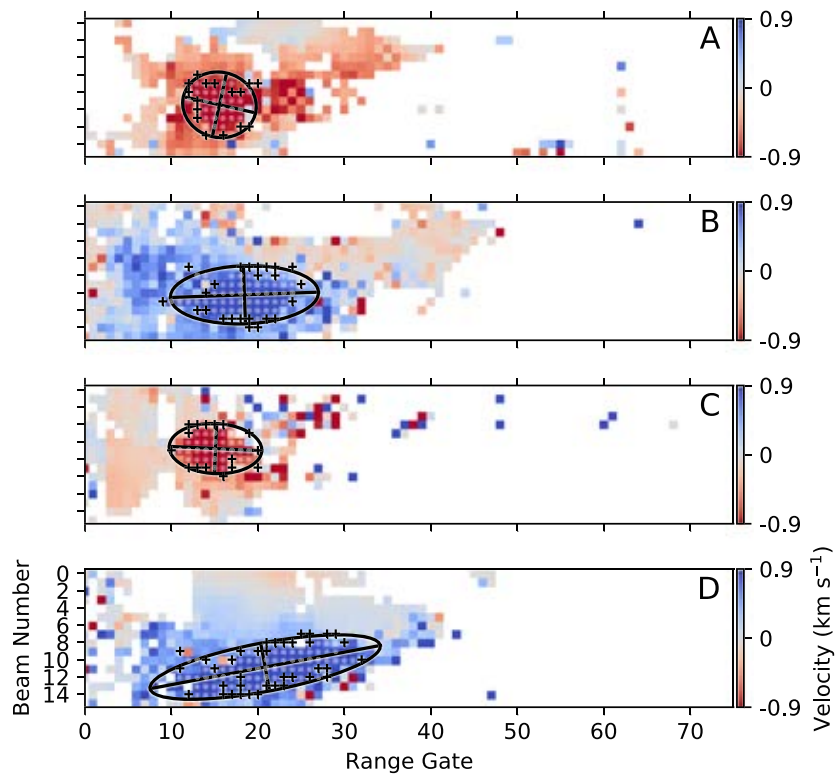

**Figure 3.** Four different examples of scans where flow channels were identified by the algorithm for the dates of (a) 28 February 2017 08:25 UT, (b) 26 October 2017 06:11 UT, (c) 15 April 2017 07:26 UT, and (d) 28 October 2017 04:33 UT.

This paper will focus on the in-depth analysis of two case studies that were identified by the detection algorithm. These two case studies are indicated by the red dots within Figure 4: Case 1 on the dawn flank and Case 2 on the dayside. Case 1 was chosen to investigate flow channels occurring deep within the polar cap, as the processes behind these channels have not yet been fully explored. Case 2 was chosen as the majority of events were detected in the dayside polar cap, so taking one from this sample allows discussion of the characteristics of flow channels on newly opened field lines. This event also persisted for almost an hour, which allows a time series to be examined and for the formation, evolution, and decay of the channel to be studied. A later paper will explore the statistics of the identified FCs, including event durations, monthly occurrence, and the IMF dependencies of FC location.

## 4. Case Studies
### 4.1. Case 1: Dawn Polar Cap Flow Channel at 79° MLAT/7 MLT

The first case study focuses on an example of a flow channel occurring on the dawn flank (79° MLAT, 7 MLT), deep within the polar cap on 2 October 2017 at 01:10 UT. Figure 5 shows solar wind data from the OMNI 1 min resolution data set, which has been time-shifted to the nose of the Earth's bow shock. Overall, the characteristics of the solar wind show average values with no strong solar wind driving ($B_z$ −0.5–2.5 nT, $B_y$ −1–2.5 nT, density ∼3.5 n/cm$^3$, velocity ∼420 km/s, and pressure ∼1.5 nPa). Figure 5 shows that during the interval of interest, there was mainly northward IMF and $B_y$ alternated between positive and negative values. The FC duration, as detected by the algorithm, is indicated by a yellow highlighted section of the plot from 01:10–01:12 UT. Flow channels occurring on the dawn and dusk flanks in the polar cap fall into the FC 2 category and happen on old-open field lines, which are field lines that have been opened on the dayside 10–30 min earlier. In the case of FC 2 on the dawn flank, a positive $B_y$ is also expected (Sandholt & Farrugia, 2012). Searching in the solar wind data for potential FC triggers that match these specifications yields the blue highlighted section between 00:45 and 00:51 UT in Figure 5. During this interval, $B_z$ takes a small magnitude southward turning to a maximum of ∼ −0.5 nT and remains negative for 6 min and $B_y$ switches from negative to positive, dominating $B_z$ with a value of +2 nT. The time delay between $B_y$ and $B_z$ and also their signs are consistent with the Sandholt and Farrugia (2009) framework for FC 2 on the dawn flank on old-open magnetic field lines.

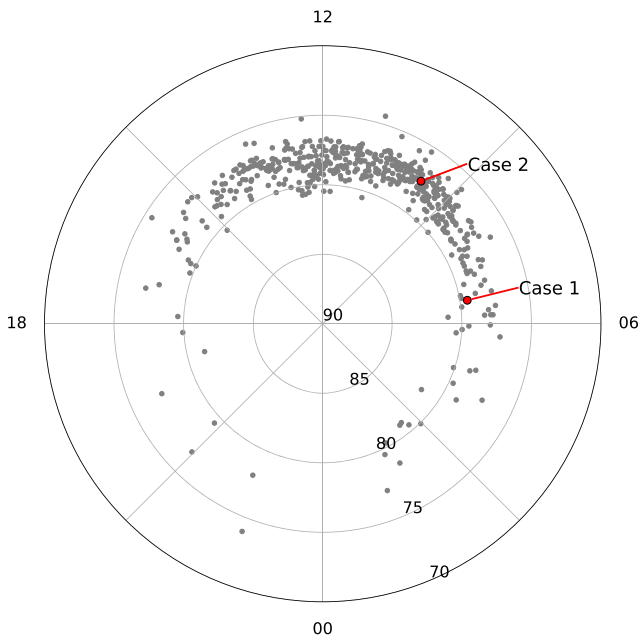

**Figure 4.** The gray dots show the occurrence in Magnetic Local Time (MLT)/Magnetic Latitude (MLat) coordinates of the flow channel center at the beginning of each event within the study. The two case studies have been indicated by red dots.

Figure 6a is a plot of data from the SSUSI onboard the F17 DMSP satellite, showing emissions in the LBHL wavelength. On the dawn flank, the UV emissions show a thick band at 70–80° MLAT associated with the auroral oval and also a thin branch further poleward (∼82° MLAT) that is aligned east-west. This feature is consistent with a PCA, specifically a bending arc. Bending arcs form under $B_y$-dominated conditions and in most studied cases $B_z$ is close to zero (Carter et al., 2015; Kullen et al., 2015). They move primarily antisunward, in contrast to other PCAs that move dawnward or duskward. The arc may have been imaged on its antisunward journey across the polar cap and was observed in the dawn sector due to the positive $B_y$ component. Figure 6b shows a SuperDARN LYR fan plot for 1:12–1:15 UT, where the FC is clearly visible as an enhanced region of antisunward flow. DMSP traversed from east to west and passed 70 MLAT on the dawnside at 01:19 UT, which means DMSP SSUSI recorded the region where the SuperDARN FOV shows the flow channel at approximately 01:12 UT, the same time as the displayed 3 min average. The FC lies in between the auroral oval and the bending arc and the flow on each side of the FC seems to slow or reverse at some ranges. The velocity shear associated with the arc is located at the poleward side of the channel. At the equatorward side, the radar begins to measure the flow reversal region associated with sunward return flow on closed field lines. The map potential plot of Figure 6c, shows a dominant dusk cell due to the positive $B_y$ and the clear signature of the flow channel in the close ranges. Figure 6c also supports the conclusion that the FC location is antisunward of the convection throat and a comparison with Figure 6a reinforces that the auroral oval emissions are in the same location as the reversal of the convection on the equatorward edge of the FC. Unfortunately, DMSP F17 did not pass directly over the flow channel, so cross-track ion drift velocity measurements are unavailable for the interval.

To investigate the temporal evolution of the flow channel and its associated optical signatures, Figure 7 (left) shows all sky camera data from Longyearbyen plotted beside SuperDARN LYR fan plots over the interval of interest (Figure 7, right). On each ASC image, north, west, south, and east are at the top, right, bottom, and left, respectively. In the first image at 01:10 UT, a stable, weak arc stretches from north-east to south-west, which has remained in place for roughly 30 min, first appearing at 00:40 UT. A second thin, faint arc formed at 1:10 UT and is visible close to this stable arc, aligned in the same direction but more directly over zenith. The arcs abruptly brighten and merge at 01:13 UT, aligning north-east to south-west. The intensification of the arcs is short-lived and begins to fade at 01:17 UT, becoming less structured and fading almost completely by 01:20 UT.

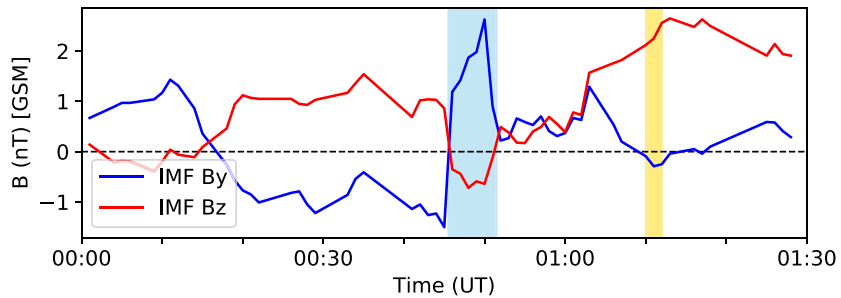

**Figure 5.** OMNI solar wind data time lagged to the bow shock for Case 1 on 2 October 2017 01:10 UT. The magnetic field data are in Geocentric Solar Magnetospheric (GSM) coordinates. The area highlighted in blue shows the potential solar wind trigger of the flow channel, while the yellow highlighted area marks the flow channel duration.

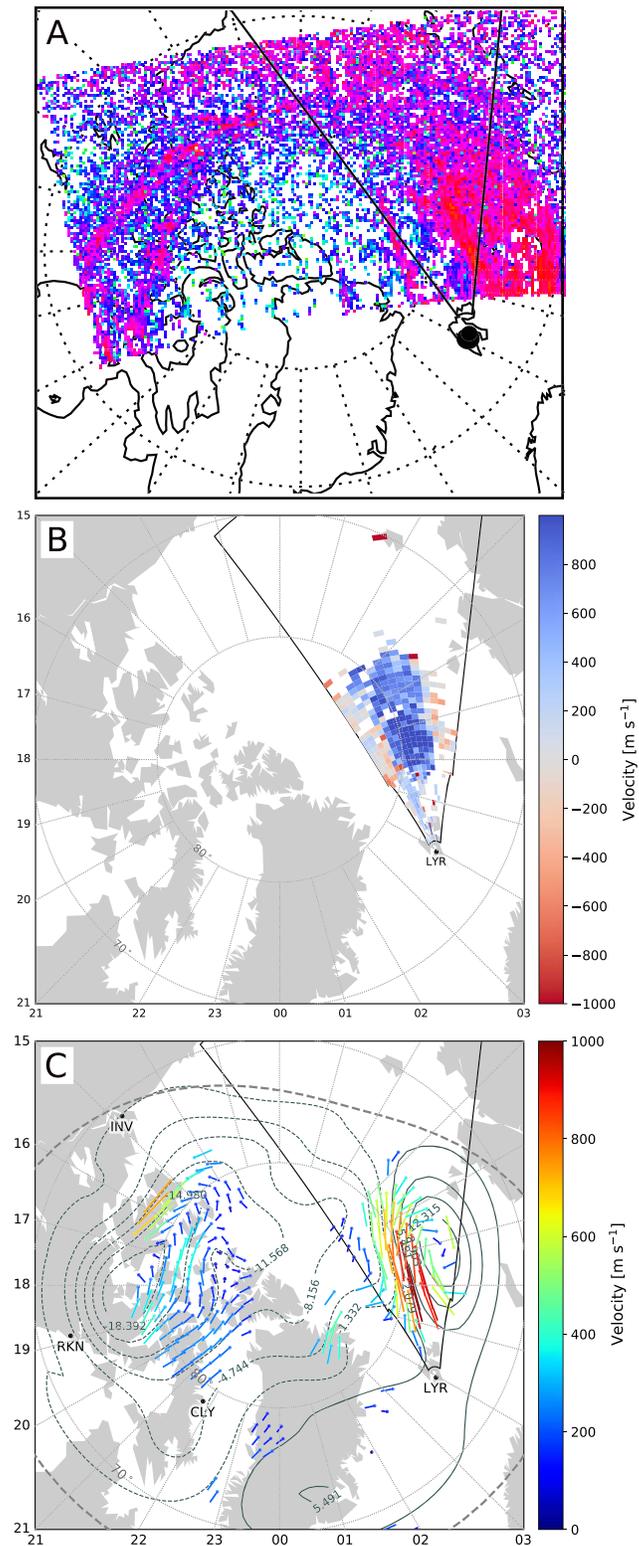

**Figure 6.** Plots for Case 1 on 2 October 2017 showing (a) the DMSP F17 pass recorded by the SSUSI instrument in the LBHL wavelength, crossing left to right over the period 1:09–1:19 UT, (b) the SuperDARN line-of-sight velocity scan from 1:12–1:15 UT, and (c) the SuperDARN convection map from 1:12–1:14 UT.

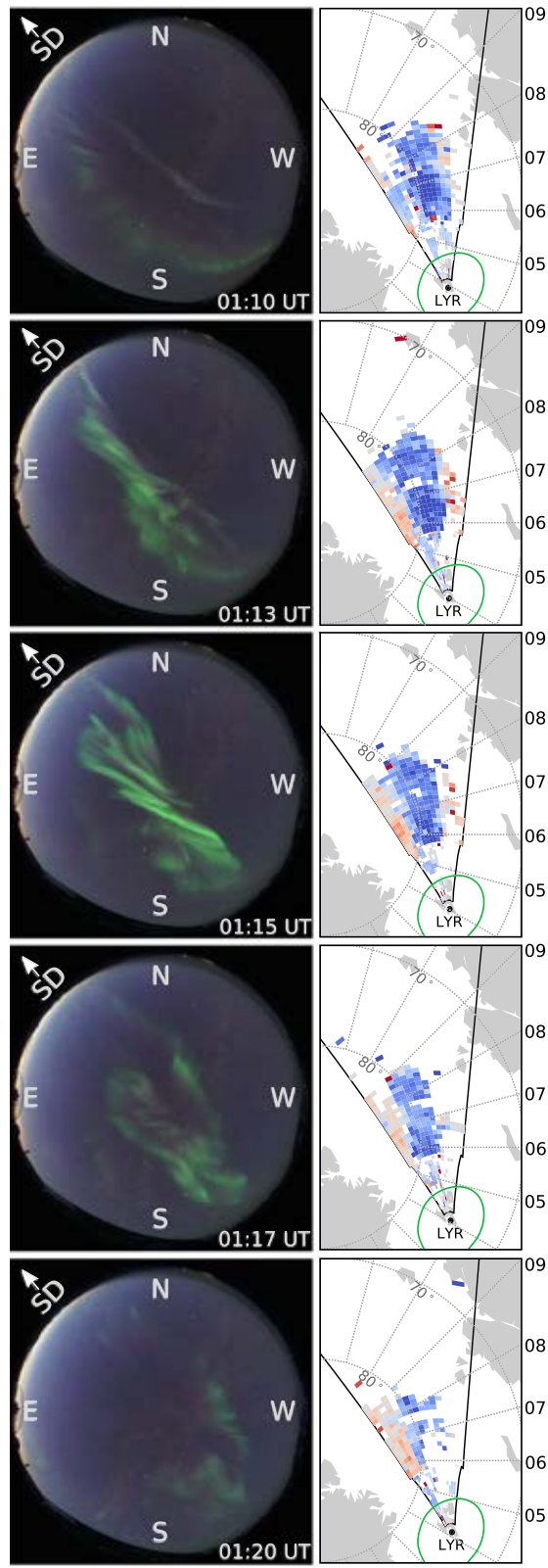

**Figure 7.** (Left column) The all sky camera images during the formation and decay of the flow channel with the cardinal points and look direction of the SuperDARN Longyearbyen radar. (Right column) The SuperDARN Longyearbyen fan plots at the same times as the all sky images, where the green oval shows an approximate field-of-view of the all sky camera at an assumed altitude of 125 km for the 557.7 nm emission.

The green oval in each panel of Figure 7 (right) shows the approximate field-of-view of the all sky camera at an assumed altitude of 125 km for the 557.7 nm emission. Figure 7 (right) shows that the FC is clearly visible at 1:10 UT. This was when the detection algorithm first identified the channel. However, the three-scan average with the greatest number of detected cells, and therefore the peak intensity of the flow channel evolution, was at 01:11 UT. There are insufficient background flows for FC detection with the algorithm by the 01:13 UT scan, although the high flows associated with the FC center still persist. A manual check of the 3 min average scans reveals that the flow channel has a duration of 13 min. The FC first begins to form at 1:07 UT, peaks at approximately 1:12 UT, and then decays in speed and size until there is very little evidence of the channel by 01:20 UT. The evolution of the FC and arc are similar in duration and intensity, which could suggest that they are a coupled ionospheric response to a system driver. The FC center is (on average over the event) located at 79° MLAT, ~7 MLT. The average velocity within the FC during the period of 01:11–01:14 UT was 985 m/s and can be used alongside the magnetic field strength to estimate an electric field value of 49 mV/m. The width as calculated from the fitted ellipse was found over the event to be on average 418 km. The potential drop over the FC can then be calculated as 21 kV. As the SuperDARN map potential value of the CPCP is 35 kV at the time of the FC, this results in the FC contributing 60% of the total CPCP.

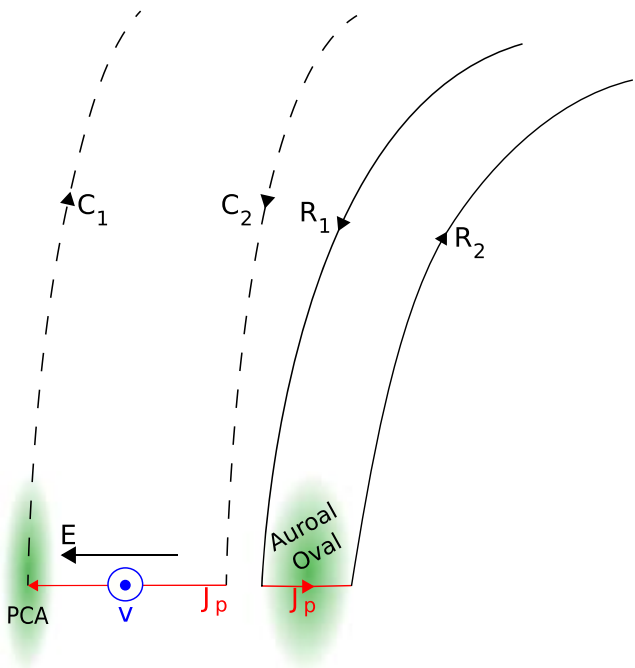

**Figure 8.** Schematic of the topology of the magnetic field, current systems, and ionospheric features for Case 1 on 2 October 2017. R1 and R2 are the Regions 1 and 2 Birkeland currents, C1 and C2 are the cusp current systems, and red arrows marked $J_p$ indicated ionospheric Pedersen currents. The flow channel velocity is depicted by a blue-filled circle, and the associated electric field by a black arrow. The polar cap arc and auroral oval are shown in green highlighted areas.

Figure 8 shows a schematic illustration of the field-aligned currents associated with the FC and PCA. On the right-hand side of the diagram are the large-scale current systems associated with the morning sector auroral oval. These currents (R1 and R2) are known as the Birkeland currents (Birkeland, 1908), which encircle the high-latitude regions of each pole in two rings (Iijima & Potemra, 1976). These current systems close via horizontal Pedersen currents, electrodynamically linking the magnetopause, the inner magnetosphere, and the ionosphere. The poleward ring (R1) flows into the ionosphere on the dawnside and out on the duskside, linking to the magnetopause current, while the polarity is reversed for R2, which maps to the partial ring current. Further poleward lie the C1-C2 cusp currents. FC 2 is the result of the closure of these currents in the ionosphere (Sandholt & Farrugia, 2009; Sandholt et al., 2006). The PCA is colocated with the upward directed C1 current. This schematic view is supported by the previously discussed flows in the fan plots shown in Figure 7 as the reversal poleward of the channel is consistent with an upward current (C1) and the auroral precipitation (PCA), while R1 is colocated with the shear between antisunward flow across the polar cap and sunward return flow on closed field lines in the dawn auroral oval.

### 4.2. Case 2: Extended Cusp Flow Channel 77° MLAT/10.5 MLT

The majority of the detected FCs occur on the dayside between 9 and 14 MLT (see Figure 4). Therefore, in order to examine a case representative of this sample, a period of multiple cusp FCs located on average at 10.5 MLT (between 05 and 06 UT) is analyzed in the following case study. Figure 9 shows the IMF magnetic field, density, velocity, and pressure from the OMNI data between 4:00 and 6:30 UT on 7 November 2017. The algorithm detected FCs in six intervals indicated by the yellow highlighted sections on all four panels in Figure 9. Figures 9b and 9d show that the solar wind density and pressure were very high during the entire interval, although Figure 9c shows a slower than average solar wind velocity. Figure 9a shows a discontinuity in the solar wind magnetic field at 4:10 UT (first black vertical line), where there is a 20 nT drop in $B_z$ to −10 nT, although there is a data gap directly before and therefore the drop could have occurred between 4:07 and 4:10 UT. At the same instance, there is a smaller drop in $B_y$ to −4 nT. By 4:45 UT, $B_y$ has gradually increased and stabilized at +10 nT. Although this period of $-B_z$ is favorable for dayside reconnection, it is probably not steady as $B_z$ is fluctuating. At 5 UT (second black line), $B_z$ turns and remains constantly southward, at which point FCs begin to be detected by the algorithm. Over the next hour, there is strong solar wind driving with a constant $+B_y$ of 10 nT and a steady $-B_z$ between −5 and −7 nT. At 5:45 UT, the density and pressure drop to half of the previous values, and $B_z$ turns positive at approximately 6:15 UT.

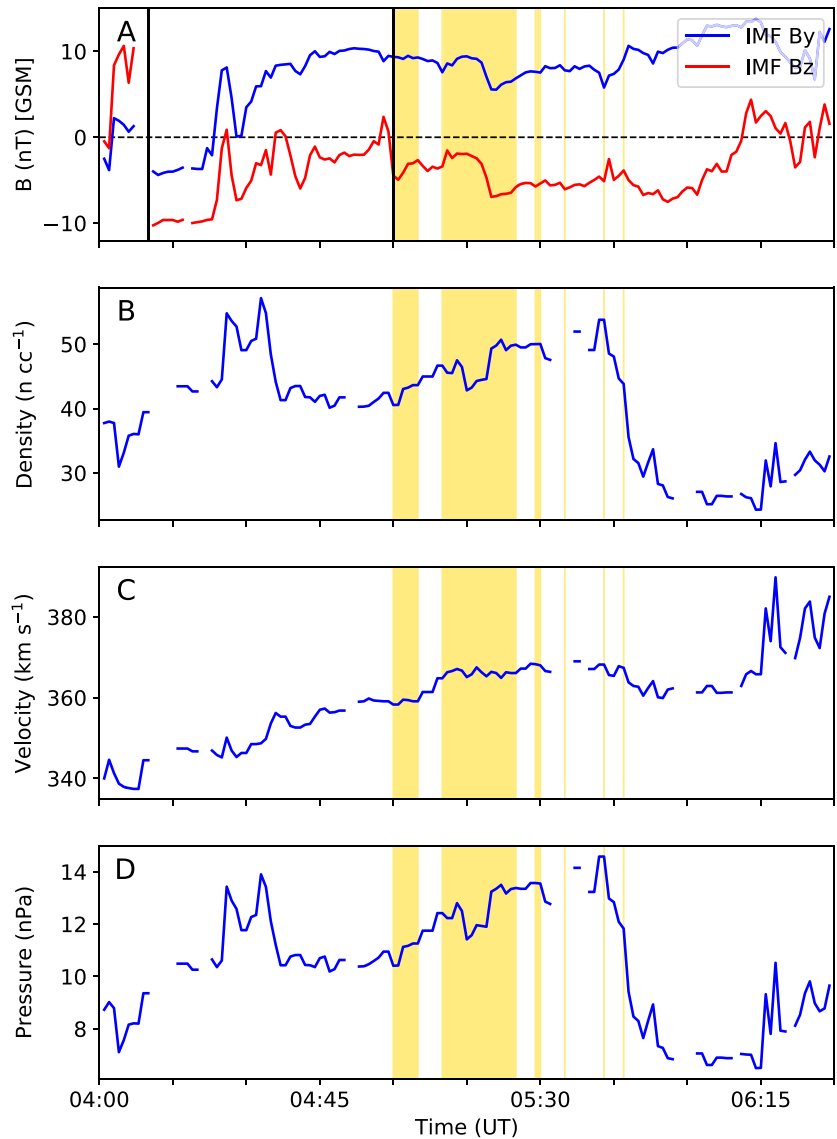

**Figure 9.** OMNI solar wind data time lagged to the bow shock for Case 2 on 7 November 2017 showing (a) IMF strength where the blue and red lines are the $B_y$ and $B_z$ components, respectively (b) density (c) velocity and (d) pressure. Yellow highlighted areas show the duration of each flow channel and the solid black lines in panel (a) mark times of interest that are elaborated on in the text.

Figure 10a shows a SuperDARN convection map where the FC was very fast and wide. The dusk cell dominates over the dawn cell, which is consistent with a positive $B_y$. Two regions of enhanced flows can be seen in this plot, a clear FC in the close ranges of the radar (77° MLAT, 10.5 MLT) and another area of >0.9 km/s flows postnoon (75° MLAT, 15 MLT). Figure 10d shows selected fan plots over the course of the event, which will be used in conjunction with the solar wind data to discuss the case study. Visual inspection of the fan plots in the lead up to the first successful FC detection reveals that the FC begins forming in the near range gates at 04:20 UT (not shown), a few minutes after the southward turning of $B_z$ at ~4:10 UT. The area of enhanced flows shows structuring, but is far below the detection threshold of the algorithm. At 4:45 UT, enhanced flows are visible in the far ranges of the SuperDARN LYR fan plot. At 5:00 UT, the far range flows are no longer visible, the FC begins to gain speed and structure in the close range gates, the velocity surpasses the threshold, and the algorithm detects the FC. The FC then persists in the close ranges until 5:47 UT.

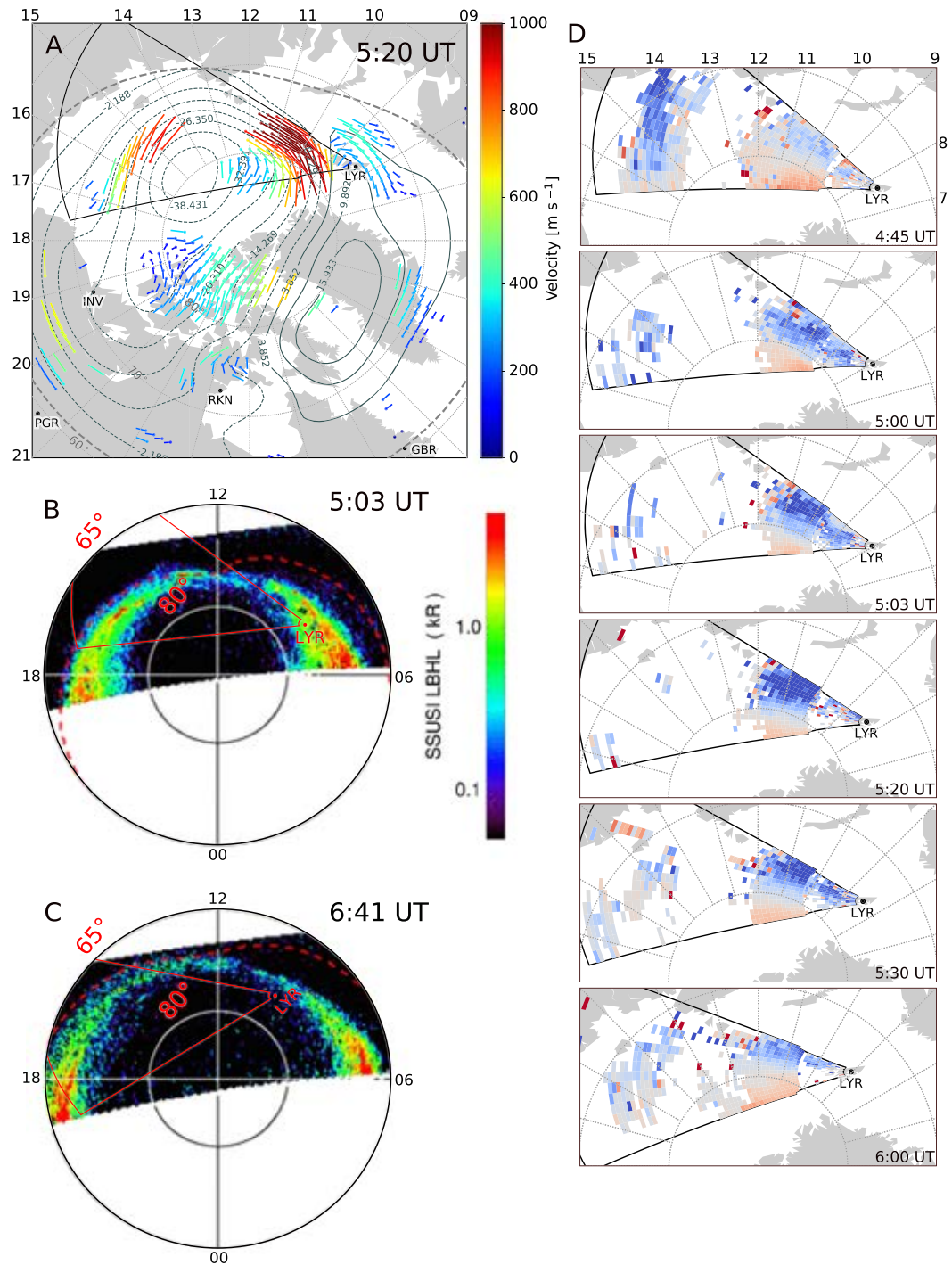

**Figure 10.** Panels for Case 2 on 7 November 2017 showing (a) the SuperDARN convection map with the LYR fan overlaid, the DMSP SSUSI data in the LBHL wavelength for (b) 5:03 UT and (c) 6:41 UT, and (d) fan plots of line-of-sight velocity over the duration of the event from the LYR radar.

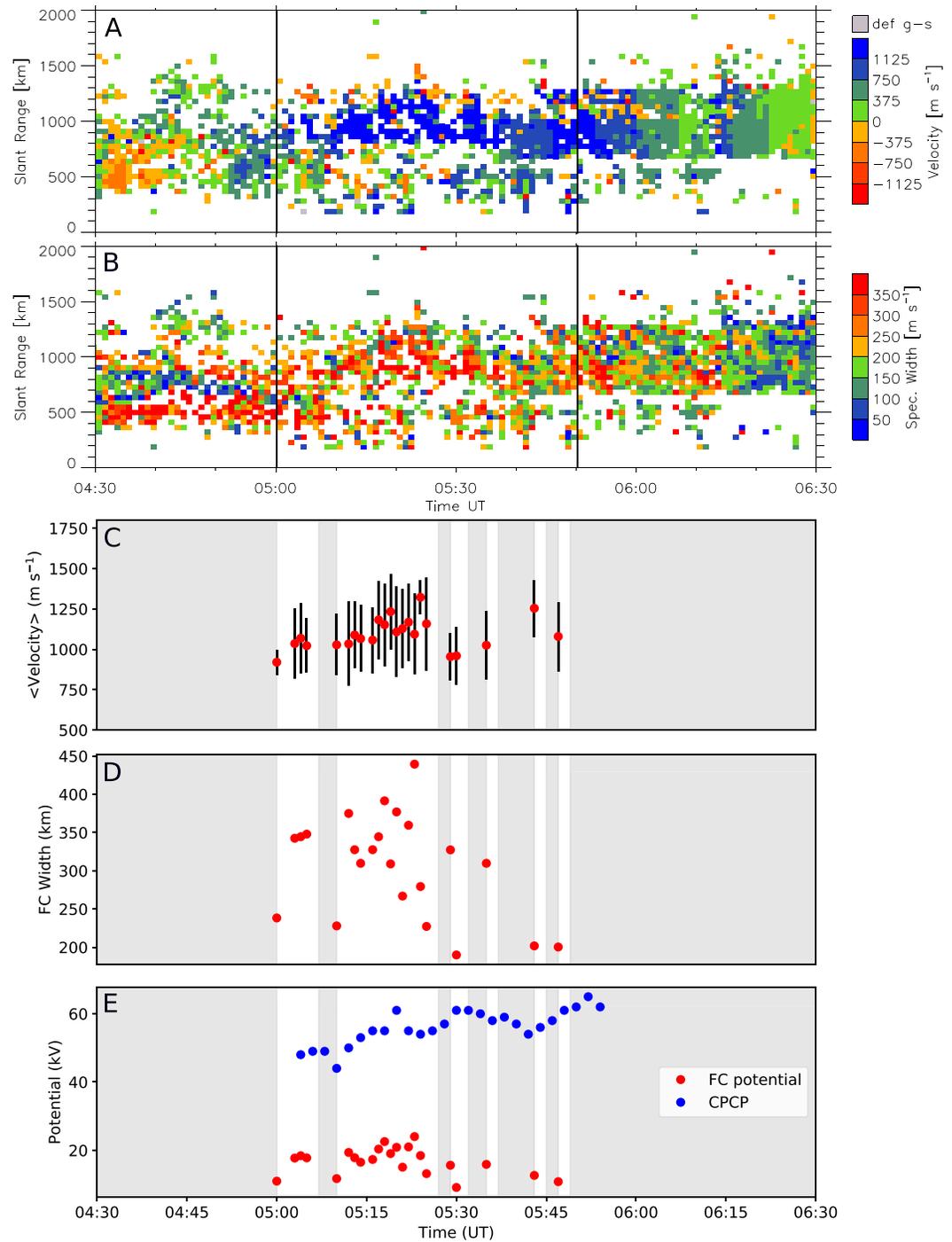

**Figure 11.** A time series of Case 2 on 7 November 2017. Panels (a) and (b) show range-time graphs for velocity and spectral width, respectively, for beam 13, in the center of the flow channel. (c) The average velocity of the flow channel (red points) with an error bar of 1 standard deviation. (d) The width of the flow channel, and (e) the flow channel potential (red points) and total cross polar cap potential (blue points).

The location of the FC within the cusp region suggests that we are observing FC 1 on newly opened field lines. The flow channel can be seen to be intermittently excited by the $-B_z$, so therefore falls under the "spontaneously driven" category. $B_z$ remains negative and drives reconnection until 6:15 UT. So rather than the FC ceasing, the disappearance of the FC from the SuperDARN LYR FOV could be due to the reconnection site moving equatorward due to the enhanced dayside reconnection. This inference is supported by the expansion of the auroral oval observed by DMSP on two separate passes as seen in Figure 10. The pass at 5:03 UT (Figure 10b) shows that at the beginning of the interval, the dayside oval sits at ∼78° MLAT, while at the next pass at 06:41 UT (Figure 10c) the oval has visibly expanded equatorward to approximately 75° MLAT, out of the range of the SuperDARN LYR radar FOV.

Figure 11 shows a time series over the course of the event, beginning 1.5 hr before the algorithm detected the FC and ending 45 min after the last detection. Panels (a) and (b) show range-time plots for beam 13, color coded by velocity and spectral width, respectively. Beam 13 was at the center of the FC, so it is ideal for studying its formation and evolution. The vertical black lines mark the first and last detection times of the algorithm during this interval. Figure 11a shows that the algorithm detected the majority of the event effectively as the highest velocity flows begin and end close to the black vertical lines. Figure 11b shows a high spectral width of up to 400 m/s inside the channel, suggesting turbulent flows. There is an apparent pulsing present in both the velocity and spectral width measurements between 5:00 and 5:35 UT, where the channel periodically moves from 800–1,200 km slant range. Figure 11c shows an average velocity (red points) at times where the algorithm detected a FC, with an error bar of 1 standard deviation to indicate the spread of velocities inside the channel. Values lower than the velocity threshold are present in the velocity spread. This is because the velocity threshold is applied to the average velocity around a cell and between the edges of the channel. This allows for significant variation in the FC at times, but the average velocity always remains above the threshold of 0.9 km/s. Figure 11c shows that the average velocity generally increases until 5:24 UT from values of 900 to 1,300 m/s. After this point there are too few points to determine a trend. Figure 11d shows a large variation in the FC width over the interval from 190–440 km, which classifies the flow channel as a mesoscale feature. Figure 11e shows the FC potential (red points) and the cross polar cap potential (blue points). The FC potential is 30% of the total cross polar cap potential on average and 46% at the peak of the event (5:23 UT). The values in Figures 11c–11e are derived from the algorithm outputs (apart from the CPCP, which is a SuperDARN data product) and the grayed-out regions on these panels indicate periods when no FCs were detected by the algorithm.

## 5. Discussion

The newly developed algorithm described in this paper identifies structured fast flows (over 0.9 km/s) within the polar cap. These flows are embedded within a slower background convection flow, with a velocity gradient of at least 400 m s$^{-1}$ cell$^{-1}$ on either side of the flow channel. Our algorithm is the first automated method for detecting FCs in SuperDARN data over all MLTs within the polar cap. The algorithm identified 546 events in the Longyearbyen radar data during 2017. The stringent criteria applied with the algorithm ensures that only well-defined, fast flow channels were identified, which are moving in the same direction as the background convection. The algorithm was most effective in detecting the FCs at their peak, at the time of highest flow velocity due to the high-velocity selection criteria. Both short and long duration events were identified at a range of MLTs. Detected FCs should be located in the polar cap in a statistical sense, as the most poleward section of the open-closed field line boundary (dayside cusp region) is statistically located at 75° latitude (Yeoman et al., 2002) and the SuperDARN LYR radar measures a further poleward latitudinal range of approximately 76–82°.

Two case studies were chosen for further investigation: Case 1 on the dawn flank and Case 2 in the cusp region. The Case Study 1 event was chosen for further discussion as relatively little is known about flow channels occurring deep within the polar cap on field lines which have been opened 10–30 min previously. Case Study 2 was chosen as the vast majority of the detected events within the Longyearbyen radar FOV occurred on the dayside. The event was interesting as it allowed discussion of a dayside event (which make up the majority of our sample) but was also chosen as it was roughly an hour in duration. This allows a meaningful time series to be examined and for the formation, evolution, and decay of the channel to be discussed. In both case studies, DMSP observations place the auroral oval equatorward of the FC observations, confirming that the FCs occur within the polar cap.

### 5.1. Key Properties of the Flow Channels

Case Study 1 on 2 October 2017 shows a short-lived flow channel, lasting 13 min by eye and detected by the algorithm for 2 min, occurring on the dawn flank. The average velocity of the flow channel was 985 m/s with an average width of 418 km. This yields an electric field value and potential drop across the FC of 49 mV/m and 21 kV, respectively.

Case Study 2 on 7 November 2017 is an example of a longer-lived flow channel in the cusp, lasting for approximately 1 hr. The FC had an average speed of 1.1 km/s, an electric field value of 55 mV/m, and an average width of 307 km. The spectral width measurements inside the channel show high values of 400 m/s, suggesting turbulent flows. The average potential across the channel is 17 kV with a peak of 25 kV.

### 5.2. The Contribution of Flow Channels to the Cross Polar Cap Potential

At their peak values, the flow channels accounted for 60% and 46% of the CPCP in Cases 1 and 2, respectively. These are higher percentages than previously observed for FC 2, for example, Andalsvik et al. (2011) (35%) and Sandholt et al. (2010) (25%). Our algorithm gives a more accurate estimate as both previously published values were observed using DMSP passes, which only give measurements over the satellite trajectory at a given instance. Our algorithm evaluates the FC in two dimensions (range and beam), sampling the radar FOV at a 1 min resolution. This allows continuous observation of the channel for as long as it remains within the radar FOV and sufficient backscatter is present. It is therefore possible to observe the FC over time and obtain average values of the potential. Also, this study does not limit the data to extreme IMF conditions, such as the interplanetary coronal mass ejections typically used by Andalsvik et al. (2011) and Sandholt et al. (2010). Case 1 shows that FCs are occurring for more average values of IMF and an unremarkable, small magnitude magnetic field can still generate FCs which account for 60% of the CPCP. FCs have high velocities and potentials but are small in geographic area, and will not reproduce as well as large-scale features in convection map contours. This is due to filtering by the finite spherical harmonic expansion and due to the influence from the map potential model. Identifying FCs in the data from the individual radars is therefore essential to detecting the smaller-scale FCs, and the LYR radar is in an optimal position for the detection of polar cap FCs. As polar cap FCs can account for such a significant fraction of the CPCP, it is important to have radar coverage in the polar cap in order to obtain realistic values of the CPCP. Without the polar cap radars, the CPCP would likely be severely underestimated.

In both cases, IMF $B_y$ is the dominant IMF component. Under these conditions, a magnetic tension within the dawn-dusk direction is applied to the newly opened magnetic field lines. The entire convection pattern reconfigures on a scale of minutes with a dominant dusk/dawn cell for positive/negative IMF $B_y$ in the Northern Hemisphere (Grocott & Milan, 2014). Reconnection with a $B_y$ component then introduces asymmetric loading of magnetic flux into the magnetospheric lobes. As FC comprise a large fraction of the total CPCP, they are efficient at transmitting this asymmetric loading into the ionospheric convection pattern. This asymmetry can then be reduced when tail reconnection occurs, for example, during substorms, and the magnetospheric lobes are asymmetrically unloaded (Ohma et al., 2018; Reistad et al., 2018).

### 5.3. Flow Channel Formation and Decay Processes

The algorithm works well at picking up the peak of the FC but does not detect the formation and decay when the velocities are below the detection threshold. The case studies were manually inspected after detection to give insight into the formation and decay processes of the FC. The FC in Case 1 is linked to a PCA, which is likely a bending arc due to the preceding solar wind conditions. The optical emissions potentially associated with the arc and FC are observed in the KHO ASC, aligned in the direction of the SD FOV. Due to a lack of overlapping fields of view between the instruments used in this case study and the different observation parameters (wavelengths and scale sizes), it is not possible to conclusively link the bending arc to the auroral features seen in the ASC. However, the features could be an ionospheric response caused by the same magnetospheric driver due to their similar orientation, duration, and modulation of intensity. These observations support previous work which find FCs occurring on the edges of PCAs (Gabrielse et al., 2018; Zou et al., 2015b).

Case 2 shows a FC intermittently excited around the cusp region, strongly driven by a dense, high-pressure IMF, high magnitudes of $+B_y$ and a sustained $-B_y$. The $-B_z$ persists for over an hour and the FCs are on newly opened field lines, which makes Case 2 a spontaneously driven FC 1 in the Sandholt framework (Sandholt et al., 2010). There are signs of the structure of the FC forming approximately half an hour before the algorithm detects the FC. The velocity however shows a rapid onset in the beam directed along the center of the FC, where speeds jump above the 0.9 km/s threshold and the FC emerges rapidly from the slower flow. Figure 11a shows possible pulsing of the velocity (5:00–5:35 UT) as the channel appears to move slightly between 800 and 1,200 km in slant range. This is expected from dayside reconnection phenomena, such as PIFs (Provan et al., 1998), but will not be further analyzed as it is outside the scope of this paper. After this period, the FC seems to stabilize in position from 5:35–6:00 UT and remain at 700–1,100 km slant range. Over the lifetime of the FC, there are high spectral widths of 400 m/s, suggesting turbulent flows within the channel, which could be structuring at smaller scales than one SuperDARN range gate (45 km).

### 5.4. Momentum Transfer on Old Open Flux

The FC in Case 1, residing on the dawn flank, was 418 km in width, lasted for 13 min (detected by the algorithm at its peak for 2 min) and was detected 25 min after a small deviation from northward to southward IMF initiated a reconnection burst with a dominant IMF $B_y$ positive component. This width, duration, and delay time indicate a directly driven FC 2 category on old open flux within the Sandholt and Farrugia (2009) framework. Despite small IMF magnitudes, fast flows are driven deep inside the polar cap, accounting for 60% of the CPCP. The FC is observed between a thin, poleward band of emissions and equatorward auroral oval emissions in the DMSP SSUSI data. Due to the $B_y$ dominant conditions and as $B_z$ is close to zero, the band of emissions most likely falls into the bending arc subclass of PCAs (Carter et al., 2015; Kullen et al., 2015). Bending arcs are located on open field lines (Carter et al., 2015), which further supports the theory that the flow channel is occurring on old open field lines on the dawn flank, deep within the polar cap. This work builds upon the work of Sandholt and Farrugia (2009), using a similar velocity threshold (0.9 km/s) as compared to the 1 km/s velocity threshold of Andalsvik et al. (2011). For the first time, FC 2 is found in conjunction with a PCA (specifically, a bending arc) on the dawn flank through combined observations from DMSP, SuperDARN LYR fan plots, and SuperDARN convection maps.

## 6. Conclusions

A new algorithm was developed to locate flow channels within the polar cap. The algorithm detected 546 events over a years' worth of data (2017) from the Longyearbyen SuperDARN radar. Two case studies were selected for further analysis: Case 1 on the dawn flank and Case 2 in the cusp. The main findings from these case studies can be summarized as follows:

- The algorithm identified 546 events within the 1 year interval. FCs were observed in the polar cap over most magnetic local times, but the vast majority were detected in the dayside polar cap.
- The FCs comprise high values of the CPCP, peaking at 60% and 46% for Cases 1 and 2, respectively. Thus, polar cap FCs play an important role in flux transfer through the solar wind-magnetosphere-ionosphere system.
- Case 1 confirms that FCs do occur on the edge of PCAs and that fast ionospheric flows can form deep inside the polar cap under small magnitude IMF fields that are $B_y$ dominant.
- Case 2 shows that fast flows can be driven in the cusp for extended periods with a negative $B_z$ component of the IMF and a high magnitude positive IMF $B_y$. The flow inside these channels is turbulent, exhibiting higher spectral widths for faster flows, which suggests structuring at spatial scales less than one SuperDARN range gate (45 km).

The two case studies provide confidence in the ability of the algorithm to identify FCs in the polar cap. A future publication will detail the findings of a statistical study using all 546 identified events.

**Acknowledgments**
We would like to thank Steve Milan, Jenny Carter, Jade Reidy, and the members of the Birkeland Centre for Space Science for useful science discussions. Financial support was provided by the Research Council of Norway under Contract 223252. Solar wind and IMF data are available at the Goddard Space Flight Center Space Physics Data Facility (https://cdaweb.sci.gsfc.nasa.gov/index.html/). SuperDARN data were obtained through the SuperDARN website at Virginia Polytechnic Institute and State University (http://vt.superdarn.org/). DMSP SSUSI data are freely available at Johns Hopkins Applied Physics Laboratory (https://ssusi.jhuapl.edu/). We thank Larry Paxton (PI of the DMSP SSUSI instrument) and Dag Lorentzen (PI of SuperDARN Longyearbyen) for use of the data. Quick look all sky camera data from the Kjell Henriksen Observatory are available online (http://kho.unis.no/) and the individual all sky camera images used in this paper are available online (https://figshare.com/articles/All_Sky_Camera_Images_Svalbard_2_2017/9971246) and were obtained from the instrument PI Dag Lorentzen (dag.lorentzen@unis.no).